\documentclass[twocolumn]{aastex6}

\newcommand{\NH}{N_\mathrm{H}}

\usepackage{graphicx}
\usepackage{CJK}
\shorttitle{Diagnosing the kinematics of the tori in AGNs} \shortauthors{Liu \& Li}

\begin{document}
\begin{CJK*}{GB}{gbsn}

\title{Diagnosing the kinematics of the tori in active galactic nuclei with the velocity-resolved reverberation mapping of the narrow iron K$\alpha$ line}


\author{Yuan Liu (ÁõÔª) and Xiaobo Li (ÀîС²¨)}

\affil{Key Laboratory of Particle Astrophysics, Institute of High
Energy Physics, Chinese Academy of Sciences, P.O.Box 918-3, Beijing
100049, China}

\email{liuyuan@ihep.ac.cn; lixb@ihep.ac.cn}

\begin{abstract}
The properties of the dusty tori in active galactic nuclei (AGNs) have been investigated in detail, mainly focusing on the geometry and components; however, the kinematics of the torus is still not clear. The narrow iron K$\alpha$ line at 6.4 keV is thought to be produced by the X-ray reflection from the torus. Thus,  the velocity-resolved reverberation mapping of it is able to constrain the kinematics of the torus. Such effort is limited by the spectral resolution of current CCD detectors and should be possible with the microcalorimeter on the next generation X-ray satellite. In this paper, we first construct the response functions of the torus under a uniform inflow, a Keplerian rotation, and a uniform outflow. Then the energy-dependent light curve of the narrow iron K$\alpha$ line is simulated according to the performance of the X-ray Integral Field Unit in Athena. Finally, the energy-dependent cross-correlation function is calculated to reveal the kinematic signal. According to our results, one hundred observations with 5 ks exposure of each are sufficient to distinguish the above three velocity fields. Although the real geometry and velocity field of the torus could be more complex than we assumed, the present result proves the feasibility of the velocity-resolved reverberation mapping of the narrow iron K$\alpha$ line. The combination of  the dynamics of the torus with those of the broad line region and the host galaxy is instructive for the understanding of the feeding and feedback process of AGNs.
\end{abstract}



\section{Introduction}

A dusty torus is proposed to unify the apparent difference between type 1
and 2 active galactic nuclei (AGNs), i.e., they are intrinsically the same but only viewed from different inclinations (Antonucci 1993; Urry \& Padovani
1995). As the  key ingredient of the unification model, it is important to understand the properties of tori, which will also shed light on the feeding and feedback process of the central supermassive black hole. There is already a considerable body of work discussing the geometry of the torus by infrared imaging, interferometry, and spectral energy distribution fitting.  Asmus et al. (2016) investigated the extended mid-infrared emission of 21 AGNs and compared it with the position angle of the system axis. They found that the mid-infrared emission is dominated by dust in the polar region. Tristram et al. (2014) constrained the geometry of the dusty torus in the Circinus galaxy  by infrared interferometry and fitted the result by a disk-like and an extended component (roughly perpendicular to the disk component). Mart\'{\i}nez-Paredes et al. (2017) fitted the spectral energy distributions of 20 nearby quasars with a clumpy torus model and found that the parameters of quasars are significantly different from those of Seyfert galaxies, i.e., a lower number of clouds, steeper radial distributions of clouds, and clouds that are less optically thick than in Seyfert 1. Despite the progress on the understanding of the geometry of tori, there is still little knowledge about the kinematics of tori. Integral field spectroscopy (IFS) can reveal the kinematics of host galaxies and the kinematics of the broad line region (BLR) is constrained by the velocity-resolved  reverberation mapping of broad emission lines in optical band (Denney et al. 2009; Bentz et al. 2008, 2009, 2010; Pancoast et al. 2011, 2012, 2014; Du et al. 2016; Grier et al. 2013, 2017); however, there is no appropriate technique for the torus, as the ``bridge'' between host galaxy and supermassive black hole. There are diverse requirements from theoretical models on the dynamics of tori. Both outflow and inflow model of tori have been proposed (Wang et al. 2010; Czerny \& Hryniewicz 2011).

The iron K$\alpha$ line at 6.4 keV in the X-ray spectra of AGNs is a valuable feature. Reverberation mapping of the broad component (FWHM$>200$ eV) can probe the geometry and dynamics of the inner accretion flow (see Uttley et al. [2014] for a comprehensive review). Kara et al. (2016) presented X-ray reverberation of all variable Seyfert galaxies with \textit{XMM-Newton} data. Velocity-resolved reverberation mapping of the broad component has also been attempted (Cackett et al. 2014).
The narrow iron K$\alpha$ line (FWHM$\sim50$ eV) is quite universal and believed to be originated from the torus. Reverberation mapping of  the narrow iron K$\alpha$ line can constrain its emitting region (Liu et al. 2010). Although it can also probe the kinematics of tori, the spectral resolution of X-ray CCD ($\sim150$ eV) is not enough to resolve  different velocity parts of the narrow iron K$\alpha$ line. Even with the \textit{Chandra} high-energy grating (the spectral resolution is $\sim40$ eV at 6.4 keV) it is hard to perform a velocity-resolved campaign as that of H$\beta$ line by optical spectroscopy. Such observations are only realistic by a microcalorimeter with spectral resolution better than 10 eV at 6.4 keV. In this paper, we will investigate the ability of microcalorimeters in the velocity-resolved reverberation mapping of the narrow iron K$\alpha$ line.

There are already some X-ray spectral models of the torus in AGN (Ikeda et al. 2009; Murphy \& Yaqoob 2009; Brightman \& Nandra 2011; Furui et al. 2016); however, they cannot provide the temporal response of the torus, which is necessary for the simulation of the reverberation mapping. We have constructed an X-ray spectral model of tori using Geant4 (Liu \& Li 2014, 2015, 2016) and the same code can also record the travelling time of every photon (including the fluorescent line).\footnote{Our X-ray spectral model is available at https://heasarc.gsfc.nasa.gov/xanadu/xspec/models/Ctorus.html} Thus, it is possible to construct the two-dimensional response (temporal and spectral) of the torus and further simulate the ability of microcalorimeters in diagnosing the kinematics of the torus. In Section 2, we will explain the construction of  the two-dimensional response. In Section 3, the light curve of NGC 5548 is adopted as the incident source and the observed spectra are simulated under the response of the X-ray Integral Field Unit (X-IFU) in Athena (Barret et al. 2016). In Section 4, we compare the results under different kinematics and discuss their implications.

\section{The Construction of response functions}
\begin{figure}[htbp]
\plotone{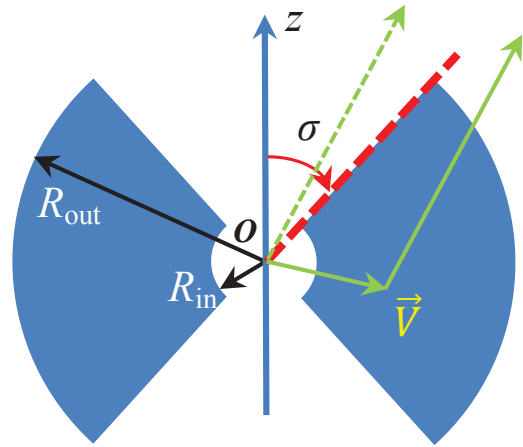}
 \caption{Cross-section view of the geometry of the smooth torus adopted in our simulations. The boundary is defined by the inner radius $R_\mathrm{in}=0.1$ pc, the
 outer radius $R_\mathrm{out}=2.0$ pc, and the half-opening angle $\sigma=60^\circ$. The gas is uniformly
distributed in the torus.
 The central X-ray source is located at $O$. The delay time of a photon with the solid green trajectory is calculated relative to the dashed green line. \label{fig1}}
\end{figure}

The spectral and temporal simulations were performed by an object-oriented toolkit, Geant4 (Agostinelli et al. 2003; version
4.9.4).\footnote{http://geant4.cern.ch/} We adopted the geometry of the smooth torus in Liu \& Li (2014; Figure \ref{fig1}).
The real structure of the torus could be more complex than this simple geometry; however, the main purpose here is to verify the
feasibility of velocity-resolved reverberation mapping of the Fe K$\alpha$ line. The simulation code is able to include more realistic structure according to specific requirements in the future.

The abundances of elements were from Anders \&
Grevesse (1989). All atoms were assumed in their ground states.
The low-energy electromagnetic process in Geant4 was invoked. More specifically, we considered the photoelectric effect (fluorescence and Auger
processes were also loaded), Compton scattering, Rayleigh
scattering, and $\gamma$ conversion for photons. Ionization, bremsstrahlung, and multiple scattering were
considered for electrons. The relevant cross sections and atomic data
were adopted from
EPDL97.\footnote{https://www-nds.iaea.org/epdl97/}

In the simulations, Geant4 tracks the trajectories of primary particles
and secondary particles step by step and  records the information of particles,
e.g., kinetic energy, momentum, position, time, and physical process
involved.  Please see Liu \& Li (2014) for more details about the simulation in Geant4.
In the simulations of this paper, the column density $\NH$  in the radial direction of the torus is $10^{24}$ cm$^{-2}$. The optical depth of the Fe K$\alpha$ line is $\sim1$ at $\NH=4\times10^{23}$ cm$^{-2}$. Thus, the Fe K$\alpha$ line can probe a considerable part of the torus with $\NH=10^{24}$ cm$^{-2}$.

The incident photons were isotropically emitted from the accretion disk/corona (the point $O$  in Figure \ref{fig1}) and their spectrum
 was assumed to be a single power law, i.e.,
$\mathrm{flux} \propto E^{-\Gamma}$ (1 keV $\leq E\leq$ 500 keV),
where $\Gamma$ is the photon index and fixed at 1.8 throughout the
simulations in this paper.

The delay time of an Fe K$\alpha$ photon was defined relative to the continuum in the same direction (Figure \ref{fig1}).
Since the material in Geant4 should be static, we need to assume a velocity field of the torus $\textbf{\textit{V}}(r,\theta,\phi)$ and then combine
it with the trajectories of photons to calculate the two-dimensional response. If there was no scattering after the photoelectric effect, i.e., an Fe K$\alpha$
 photon directly escapes from the torus, the observed energy of this Fe K$\alpha$ photon (after Doppler effect) was determined by the velocity projected onto the observed direction. If there were scatterings after the photoelectric effect, for each scattering the energy of the incoming photon was adjusted according to the Doppler effect. The direction of the scattered photon was assumed to be the same as that in the static case. The energy of the scattered photon was calculated by the Compton scattering relation. The above procedure was performed on each photon; then a two-dimensional histogram was constructed on the observed energy $E$ and delay time $\tau$ plane, which is the  two-dimensional response function $RSP(E,\tau)$.

\begin{figure}
\plotone{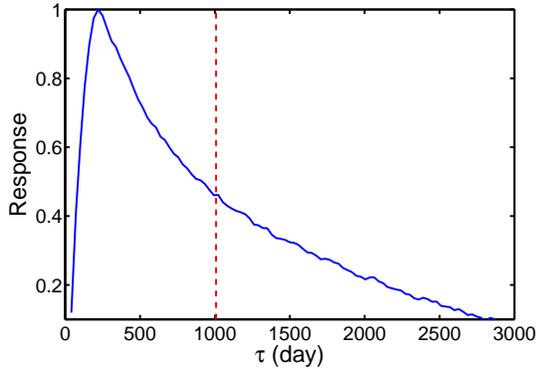}
 \caption{Response of the torus as a function of the delay time (normalized at the maximum). The red dashed line is the mean delay time (1007.9 days).\label{t_rsp}}
\end{figure}

\begin{figure*}
\plotone{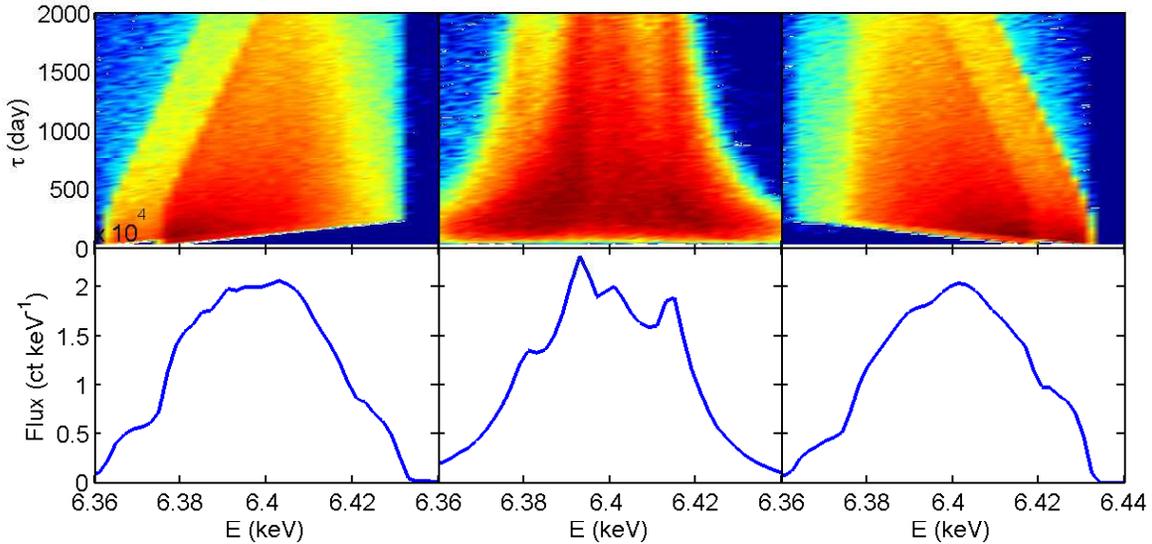}
 \caption{Upper row is the  two-dimensional response of the torus as the function of the observed energy and delay time. The strength of the response is shown in logarithmic scale. The lower row is the profile of the Fe K$\alpha$ line.  The left, middle, and right columns are the results of inflow, Keplerian rotation, and outflow, respectively. \label{rsp2d}}
\end{figure*}

 We simulated three kinds of velocity fields, i.e., a uniform inflow $\textbf{\textit{V}}(r,\theta,\phi)=-1500$ km s$^{-1}\textbf{\textit{n}}_r$, a Keplerian rotation $\textbf{\textit{V}}(r,\theta,\phi)=4000$ km s$^{-1} (r\sin\theta/0.1$ pc)$^{-1/2}\textit{\textbf{n}}_\phi$, and a uniform outflow $\textbf{\textit{V}}(r,\theta,\phi)=1500$ km s$^{-1}\textbf{\textit{n}}_r$. The FWHMs of the Fe K$\alpha$ line under the above three kinds of velocity fields were $\sim2000$ km s$^{-1}$. The  two-dimensional response functions and the profiles of the Fe K$\alpha$ line are shown in Figure \ref{rsp2d}. The response function was calculated for $\cos\theta=0.8-0.9$, i.e., type 1 AGN. Figure \ref{t_rsp} is the response as the function of the delay time, which only depends on the geometry of the torus and is thus the same for different  velocity fields.

 \section{The simulated spectra}

  \begin{figure}[b!]
 \plotone{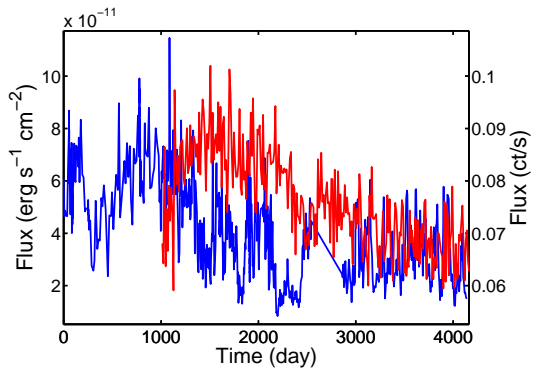}
 \caption{X-ray light curve of NGC 5548 in 2-10 keV (blue) and the simulated light curve of the narrow Fe K$\alpha$ line (red). The beginning of the light curve is MJD 50208.037326.\label{ngc5548}}
 \end{figure}

 The X-ray light curve of NGC 5548 from the \textit{Rossi X-ray Timing Explorer} in 2-10 keV was adopted as the template for the continuum variability (Figure \ref{ngc5548}; Rivers et al. 2013).
\footnote{Please see data reduction details at http://cass.ucsd.edu/$\sim$rxteagn/NGC5548/NGC5548.html} For simplicity, the photon index was assumed to be a constant (1.8)
during the variation. Sobolewska \& Papadakis (2009) showed
that the continuum slope of NGC 5548 only weakly depends on its flux.

 The light curve in Figure \ref{ngc5548} was interpolated at the interval of 11.5 days, i.e., the same temporal resolution  of $RSP(E,\tau)$, and the  interpolated light curve was then convolved with $RSP(E,\tau)$ in Figure \ref{rsp2d} to produce the spectral model of the Fe K$\alpha$ line at different times. The interval selected here is somewhat arbitrary, but the final result is not sensitive to the exact value of the interval. A simulated light curve of the integrated Fe K$\alpha$ line is shown in Figure \ref{ngc5548}. Only the light curve after the mean delay time (see Figure \ref{t_rsp}) is presented to avoid the artificial trend  due to the incompleteness of the flux of the continuum.
 To produce the final spectral model, the continua (a single power law with the flux interpolated from Figure \ref{ngc5548}) were added with the spectra of the Fe K$\alpha$ line at the corresponding time and the mean equivalent width of the Fe K$\alpha$  line was
 set to 100 eV.

 Then we adopted the above spectral model and simulated the observed spectra using the latest response and background of Athena/X-IFU.\footnote{http://x-ifu.irap.omp.eu/resources-for-users-and-x-ifu-consortium-members/}
Since the delay time of the Fe K$\alpha$ line from the torus is about several hundreds of days, long-term monitoring is required to perform the reverberation mapping. The calculation of the response of the Fe K$\alpha$ line needs the preceding flux of the continuum; thus, we first calculated the weighted average delay time of each energy bin $\bar{\tau}(E)=\int\tau RSP(E,\tau) d\tau/\int RSP(E,\tau) d\tau$ and selected 100 observations just after $\bar{\tau}(E)$ with the  interval of 23 days to reduce the artificial correlation due to the incompleteness of the flux of the continuum.
The exposure time of each observation was 5 ks and we will discuss the effect of different intervals and number of observations in Section 4.
To determine the flux of the Fe K$\alpha$ line, we fitted the simulated spectra (excluding the spectra in 6.0-6.5 keV) with a power law and then subtracted it from the total spectra to obtain the spectra of the Fe K$\alpha$ line. The original size of a spectral bin of Athena/X-IFU is 0.4 eV. We binned every 24 original bins to produce the energy-dependent light curve of the  Fe K$\alpha$ line of the spectral resolution of 9.6 eV.
Then the light curve of the continuum in 2-10 keV was cross-correlated with every bin of the  light curve of the Fe K$\alpha$ line.
 The cross-correlation function (CCF) as the function of energy and delay time is shown in Figure \ref{ccf}.

\section{Results and discussion}

 \begin{figure*}[htbp]
\plotone{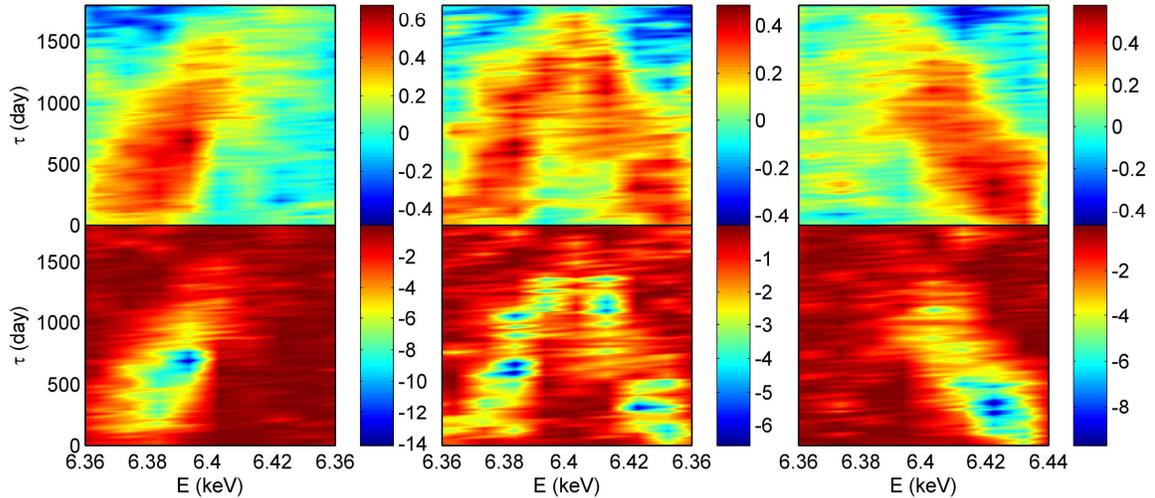}
 \caption{Upper row is the  CCF as the function of the rest-frame energy and delay time. The lower row is the  logarithm of the significance of correlations as the function of the rest-frame energy and delay time.  The left, middle, and right columns are the results of inflow, Keplerian rotation, and outflow, respectively. \label{ccf}}
\end{figure*}

As shown in Figure \ref{ccf}, it is sufficient to distinguish different kinematics of the torus by the simulated CCF, which is ensured by the large effective area ($\sim$2500 cm$^2$ at 6.4 keV) and small spectral resolution (2.5 eV) of Athena/X-IFU.
We have also performed similar simulations for Astro-H under the same exposure time and the number of observations.
The signal of correlation is weak and it is hard to distinguish different kinematics. To obtain the similar strength of the correlation, we should increase the exposure time of  each observation to 10 ks and the number of observations to 400 with the interval of 1 day, which is not realistic for an X-ray satellite.

Since the outer radius of the torus in our simulations is 2 pc, the delay between the Fe K$\alpha$ line and the continuum  can be longer than 1500 days. Thus, a long-term monitoring of about 5 years is required. However, the delay time of the inner radius of the torus is much shorter ($\sim$100 days). As shown in Figure \ref{ccf}, the pattern of the blue and red wings of the Fe K$\alpha$ line can also indicate different velocity fields. Thus, it is appropriate to first perform a pilot and short program focusing on the response of the wings of the Fe K$\alpha$ line. Nevertheless, the real velocity field can be more complex than the three cases investigated here or a combination of them. A long-term program to map the full response is always desired, of which the monitoring on the continuum can be completed by other X-ray satellites or shorter exposures of Athena. For simplicity, we adopted the  same intervals of the light curve of the continuum and the Fe K$\alpha$ line. They can be different in the real observations and a denser sampling will obviously increase the significance of the pattern in  Figure \ref{ccf}. If we increase the interval of observations to 46 days and reduce the total number of exposure to 50, the pattern in Figure \ref{ccf} can still be revealed; however, the significance of the majority of the plane will higher than 0.01. A larger spectral bin of $\sim$20 eV is necessary in this case.

In the adopted spectral model, we assumed the whole Fe K$\alpha$ line was contributed by the torus. Although the Fe K$\alpha$ lines of some AGNs are indeed dominated by the narrow component, there could be some contamination from the broad Fe K$\alpha$ line around the accretion disk in the general case. Since the response of the broad Fe K$\alpha$ line is much faster than the narrow one, it should not produce a significant signal at long delay time and thus will not contaminate the pattern in Figure \ref{ccf} if the signal-to-noise ratio (S/N) of the narrow component is enough. Besides the Fe K$\alpha$ line, the Fe K$\beta$ line (if observable) can be also utilized to perform velocity-resolved  reverberation mapping and confirm the result from the Fe K$\alpha$ line.

The amplitude of the variation of the Fe K$\alpha$ line is critical for the reverberation mapping. As shown in Figure \ref{ngc5548}, the variation of the simulated light curve of the Fe K$\alpha$ line is moderate ($F_{\textrm{var}}\sim$12\%).\footnote{$F_{\rm{var}} = \sqrt{(S^2 - \langle \sigma^2 \rangle)/ \langle X
\rangle^2}$, where $S^2$ is the variance of the whole light curve, $\langle
\sigma^2 \rangle $ is the mean square error, and $ \langle X \rangle
$ is the mean count rate of the whole light curve.}  Shu et al. (2010) compiled a sample of Seyfert galaxies observed by the Chandra high-energy grating. For the sources observed repeatedly, the difference in the best-fitting flux of the Fe K$\alpha$ line can reach the level of the variation in Figure \ref{ngc5548}; however, the significance of the difference is no more than 99\% confidence level due to limited S/N.  Fukazawa et al. (2016) systematically investigated the variability of the narrow Fe K$\alpha$ line for Seyfert galaxies using \textit{Suzaku} and \textit{XMM-Newton} data and  detected the variability of several tens of percent at $2\sigma$ level. Although long-term monitoring with high spectral resolution and S/N is still required to unambiguously determine the variability of  the narrow Fe K$\alpha$ line, it is not unrealistic to expect the sufficient amplitude of the variation to carry out reverberation mapping of the narrow Fe K$\alpha$ line.

In our simulations, the geometry of the torus does not change with the flux of the incident continuum. Koshida et al. (2009) claimed a strong variation of the inner radius of the torus in NGC 4151 by the dust reverberation mapping; however, other works indicated that the hot dust in NGC 4151 does not expand but they only detected an increase in temperature (Pott et al. 2010; H{\"o}nig \& Kishimoto 2011; Schn{\"u}lle et al. 2013). Contrary trends of the change of the lag time were obtained even in the same epoch of NGC 4151 (Kishimoto et al. 2013; Schn{\"u}lle et al. 2017). The reverberation mapping of the Fe K$\alpha$ line, especially the kinematic information, should be helpful in clarifying the discrepancy and distinguishing different models.

The velocity-resolved  reverberation mapping of BLR shows diverse signals.
Denney et al. (2009) found the results from NGC 3227, NGC 3516, and NGC 5548 show evidence for outflowing, infalling, and virialized BLR gas motions, respectively. Du et al. (2016) confirmed a similar and diverse behaviour of super-Eddington accreting AGNs. It is not clear what is the driving factor that determines the dynamics of BLR and if there is any evolutionary sequence. The dynamics of the torus is also under debate. Wang et al. (2010) required the torus as an inflow to feed the BLR and accretion disk. Czerny \& Hryniewicz (2011) thought the torus is only one part of the outflow from the accretion disk.
With the velocity-resolved  reverberation mapping of the Fe K$\alpha$ line, it is possible to compare the kinematics of the BLR and torus of the same AGN (e.g. NGC 5548), which should be instructive for the mechanism driving the dynamics of the inner part of AGNs; further combining the results of IFS in optical and maser (Masini et al. 2016; Sch{\"o}nell et al. 2017), we can build a whole pattern of the matter flow from sub-pc to kpc, which is important for the understanding of the activity of  supermassive black holes.

\acknowledgments The authors thank the referee for constructive comments. This work is supported by the National Program on Key Research and Development Project (Grant No. 2016YFA0400802), and by the National Natural Science Foundation of China under grant Nos. 11573027, 11103019, and 11133002. This work has made use of light curves provided by the University of California, San Diego Center for Astrophysics and Space Sciences, X-ray Group (R.E. Rothschild, A.G. Markowitz, E.S. Rivers, and B.A. McKim), obtained at http://cass.ucsd.edu/¡«rxteagn/.

\end{CJK*}


\begin{thebibliography}{}
\bibitem[Agostinelli et al.(2003)]{2003NIMPA.506..250A} Agostinelli, S., Allison, J., Amako, K., et al.\ 2003, Nuclear Instruments and Methods in Physics Research A, 506, 250


\bibitem[Anders \& Grevesse(1989)]{1989GeCoA..53..197A} Anders, E., \& Grevesse, N.\ 1989, \gca, 53, 197


\bibitem[Antonucci(1993)]{1993ARA&A..31..473A} Antonucci, R.\ 1993, \araa, 31, 473


\bibitem[Asmus et al.(2016)]{2016ApJ...822..109A} Asmus, D., H{\"o}nig, S.~F., \& Gandhi, P.\ 2016, \apj, 822, 109


\bibitem[Barret et al.(2016)]{2016SPIE.9905E..2FB} Barret, D., Lam Trong, T., den Herder, J.-W., et al.\ 2016, \procspie, 9905, 99052F

\bibitem[Bentz et al.(2008)]{2008ApJ...689L..21B} Bentz, M.~C., Walsh, J.~L., Barth, A.~J., et al.\ 2008, \apjl, 689, L21

\bibitem[Bentz et al.(2009)]{2009ApJ...705..199B} Bentz, M.~C., Walsh, J.~L., Barth, A.~J., et al.\ 2009, \apj, 705, 199


\bibitem[Bentz et al.(2010)]{2010ApJ...716..993B} Bentz, M.~C., Walsh, J.~L., Barth, A.~J., et al.\ 2010, \apj, 716, 993



\bibitem[Brightman \& Nandra(2011)]{2011MNRAS.413.1206B} Brightman, M., \& Nandra, K.\ 2011, \mnras, 413, 1206

\bibitem[Cackett et al.(2014)]{2014MNRAS.438.2980C} Cackett, E.~M., Zoghbi, A., Reynolds, C., et al.\ 2014, \mnras, 438, 2980

\bibitem[Czerny \& Hryniewicz(2011)]{2011A&A...525L...8C} Czerny, B., \& Hryniewicz, K.\ 2011, \aap, 525, L8

\bibitem[Denney et al.(2009)]{2009ApJ...704L..80D} Denney, K.~D., Peterson, B.~M., Pogge, R.~W., et al.\ 2009, \apjl, 704, L80

\bibitem[Denney et al.(2010)]{2010ApJ...721..715D} Denney, K.~D., Peterson, B.~M., Pogge, R.~W., et al.\ 2010, \apj, 721, 715


\bibitem[Du et al.(2016)]{2016ApJ...820...27D} Du, P., Lu, K.-X., Hu, C., et al.\ 2016, \apj, 820, 27

\bibitem[Fukazawa et al.(2016)]{2016ApJ...821...15F} Fukazawa, Y., Furui, S., Hayashi, K., et al.\ 2016, \apj, 821, 15


\bibitem[Furui et al.(2016)]{2016ApJ...818..164F} Furui, S., Fukazawa, Y., Odaka, H., et al.\ 2016, \apj, 818, 164


\bibitem[Grier et al.(2013)]{2013ApJ...764...47G} Grier, C.~J., Peterson, B.~M., Horne, K., et al.\ 2013, \apj, 764, 47

\bibitem[Grier et al.(2017)]{2017arXiv170502346G} Grier, C.~J., Pancoast, A., Barth, A.~J., et al.\ 2017, arXiv:1705.02346

\bibitem[H{\"o}nig \& Kishimoto(2011)]{2011A&A...534A.121H} H{\"o}nig, S.~F., \& Kishimoto, M.\ 2011, \aap, 534, A121


\bibitem[Ikeda et al.(2009)]{2009ApJ...692..608I} Ikeda, S., Awaki, H., \& Terashima, Y.\ 2009, \apj, 692, 608


\bibitem[Kara et al.(2016)]{2016MNRAS.462..511K} Kara, E., Alston, W.~N., Fabian, A.~C., et al.\ 2016, \mnras, 462, 511

\bibitem[Kishimoto et al.(2013)]{2013ApJ...775L..36K} Kishimoto, M., H{\"o}nig, S.~F., Antonucci, R., et al.\ 2013, \apjl, 775, L36


\bibitem[Koshida et al.(2009)]{2009ApJ...700L.109K} Koshida, S., Yoshii, Y., Kobayashi, Y., et al.\ 2009, \apjl, 700, L109


\bibitem[Liu et al.(2010)]{2010ApJ...710.1228L} Liu, Y., Elvis, M., McHardy, I.~M., et al.\ 2010, \apj, 710, 1228

\bibitem[Liu \& Li(2014)]{2014ApJ...787...52L} Liu, Y., \& Li, X.\ 2014, \apj, 787, 52

\bibitem[Liu \& Li(2015)]{2015MNRAS.448L..53L} Liu, Y., \& Li, X.\ 2015, \mnras, 448, L53

\bibitem[Liu \& Li(2016)]{2016ApJ...828..114L} Liu, Y., \& Li, X.\ 2016, \apj, 828, 114

\bibitem[Mart{\'{\i}}nez-Paredes et al.(2017)]{2017MNRAS.468....2M} Mart{\'{\i}}nez-Paredes, M., Aretxaga, I., Alonso-Herrero, A., et al.\ 2017, \mnras, 468, 2


\bibitem[Masini et al.(2016)]{2016A&A...589A..59M} Masini, A., Comastri, A., Balokovi{\'c}, M., et al.\ 2016, \aap, 589, A59


\bibitem[Murphy \& Yaqoob(2009)]{2009MNRAS.397.1549M} Murphy, K.~D., \& Yaqoob, T.\ 2009, \mnras, 397, 1549


\bibitem[Pancoast et al.(2011)]{2011ApJ...730..139P} Pancoast, A., Brewer, B.~J., \& Treu, T.\ 2011, \apj, 730, 139

\bibitem[Pancoast et al.(2012)]{2012ApJ...754...49P} Pancoast, A., Brewer, B.~J., Treu, T., et al.\ 2012, \apj, 754, 49

\bibitem[Pancoast et al.(2014)]{2014MNRAS.445.3073P} Pancoast, A., Brewer, B.~J., Treu, T., et al.\ 2014, \mnras, 445, 3073

\bibitem[Pott et al.(2010)]{2010ApJ...715..736P} Pott, J.-U., Malkan, M.~A., Elitzur, M., et al.\ 2010, \apj, 715, 736

\bibitem[Rivers et al.(2013)]{2013ApJ...772..114R} Rivers, E., Markowitz, A., \& Rothschild, R.\ 2013, \apj, 772, 114

\bibitem[Sch{\"o}nell et al.(2017)]{2017MNRAS.464.1771S} Sch{\"o}nell, A.~J., Jr., Storchi-Bergmann, T., Riffel, R.~A., \& Riffel, R.\ 2017, \mnras, 464, 1771


\bibitem[Schn{\"u}lle et al.(2013)]{2013A&A...557L..13S} Schn{\"u}lle, K., Pott, J.-U., Rix, H.-W., et al.\ 2013, \aap, 557, L13


\bibitem[Schn{\"u}lle et al.(2015)]{2015A&A...578A..57S} Schn{\"u}lle, K., Pott, J.-U., Rix, H.-W., et al.\ 2015, \aap, 578, A57


\bibitem[Shu et al.(2010)]{2010ApJS..187..581S} Shu, X.~W., Yaqoob, T., \& Wang, J.~X.\ 2010, \apjs, 187, 581

\bibitem[Tristram et al.(2014)]{2014A&A...563A..82T} Tristram, K.~R.~W., Burtscher, L., Jaffe, W., et al.\ 2014, \aap, 563, A82


\bibitem[Urry \& Padovani(1995)]{1995PASP..107..803U} Urry, C.~M., \& Padovani, P.\ 1995, \pasp, 107, 803

\bibitem[Uttley et al.(2014)]{2014A&ARv..22...72U} Uttley, P., Cackett, E.~M., Fabian, A.~C., Kara, E., \& Wilkins, D.~R.\ 2014, \aapr, 22, 72

\bibitem[Wang et al.(2010)]{2010ApJ...719L.148W} Wang, J.-M., Yan, C.-S., Gao, H.-Q., et al.\ 2010, \apjl, 719, L148
\end{thebibliography}
\end{document}